\documentclass[10pt]{JHEP3}
\usepackage{amssymb}
\usepackage{amsmath}
\usepackage{float}
\usepackage[dvips]{epsfig}

\title{Quantum Machian Time in Toy Models of Gravity}
\author{Sean Gryb$^{a,b}$\\
$^a$ Perimeter Institute for Theoretical Physics\\Waterloo, Ontario N2L 2Y5, Canada \\
$^b$ Department of Physics and Astronomy, University of Waterloo\\Waterloo, Ontario N2L 3G1, Canada \\
\\Email: \email{sgryb@perimeterinstitute.ca} }
\date{\today}
\preprint{arXiv:0810.4152}

\abstract{
General Relativity on closed spatial topologies can be derived, using a technique called \emph{best-matching}, as an evolving 3-geometry subject to constraints. These constraints can be thought of as a way of imposing temporal and spatial relationalism. The same type of constraints can be used in non-relativistic particle models to produce relational theories that suffer from the same Problem of Time as that encountered in General Relativity. As a result, these simple toy models are well suited for studying the Problem of Time in quantum gravity. In this paper, a version of these particle models is studied where we \emph{best-match} the time translational invariance of the theory. Using insights gained from this procedure, we can move back and forth between absolute and relational time by changing the way in which the relational fields are varied. We then proceed to quantize this theory using Dirac and path integral quantizations. We discover that one of the constraints of the theory, which we call the Mach constraint, is responsible for removing the dependence of the theory on a background structure. It is this Mach constraint that is responsible for making the theory temporally relational. Because of the deep relationship between these models and General Relativity, this work may shed new light on the Problem of Time in quantum gravity and how one might expect time to emerge on quantum subsystems of the universe.
}

\newcommand{\eq}[1]{Eq.~(\ref{eq:#1})}

\newcommand{\scn}[1]{Sec.~(\ref{sec:#1})}
\newcommand{\apx}[1]{Appendix~(\ref{apx:#1})}

\newcommand{\abs}[1]{\ensuremath{\left| #1 \right|}}

\newcommand{\ket}[1]{\ensuremath{\left| #1 \right>}}

\newcommand{\diby}[2]{\ensuremath{\frac{\partial #1}{\partial #2}}}
\newcommand{\equa}[1]{\begin{equation} #1 \end{equation}}
\newcommand{\pb}[2]{\ensuremath{\lf\{#1,#2 \rt\}}}

\def\lf {\ensuremath{\left}}
\def\rt {\ensuremath{\right}}
\def\ra {\ensuremath{\rightarrow}}

\def\qand {\ensuremath{\quad\text{and}}}

\def\pia {\ensuremath{\pi^\alpha}}
\def\omegaa {\ensuremath{\omega^\alpha}}
\def\ta {\ensuremath{t_\alpha}}
\def\pIt {\ensuremath{p_I^T}}
\def\qI {\ensuremath{q_I}}
\def\dotomega {\ensuremath{\dot{\omega}}}
\def\dotq {\ensuremath{\dot{q}}}
\def\dotp {\ensuremath{\dot{p}}}
\def\dota {\ensuremath{\dot{a}}}
\def\opq {\ensuremath{\hat{q}}}
\def\opp {\ensuremath{\hat{p}}}

\def\jbb {\ensuremath{\text{JBB}}}
\def\pnm {\ensuremath{\text{PPM}}}
\def\ham {\ensuremath{\mathcal{H}}}
\def\lin {\ensuremath{\mathcal{L}}}
\def\calg {\ensuremath{\mathcal{G}}}
\def\calf {\ensuremath{\mathcal{F}}}
\def\calp {\ensuremath{\mathcal{P}}}
\def\cald {\ensuremath{\mathcal{D}}}
\def\Eps {\ensuremath{\mathcal{E}}}
\def\schro {Schr\"{o}dinger }

\begin{document}

\section{Intro}

Few debates in the history of physics have shed more ink than that of relationalism versus absolutism. Pragmatists take refuge in absolute structures which provide excellent arenas for performing calculations that can accurately predict the outcomes of laboratory experiments. The epistemologically minded physicist argues for relationalism on the grounds of sound philosophical foundations. Though the debate has evolved considerably since the time of Newton and Leibniz, it is still alive and well today.

To say that Einstein's presentation of General Relativity (GR) has settled the debate, even classically, is wishful thinking. Julian Barbour makes the point bluntly: ``Einstein's approach to Mach's principle has generated much confusion.''\cite{barbour:mach_principle} The confusion that Barbour is referring to (on top of some other important issues) is the distinction between a relational theory and a generally covariant theory. Einstein was led to GR in an attempt to write the laws of gravity in a generally covariant form \cite{einstein:gen_rel}. The idea was to implement Mach's principle by finding generally covariant laws of physics. These motivations are still taught in introductory courses on GR despite the fact that less than a year after Einstein published his theory, Kretschmann \cite{kretschmann:gen_cov} demonstrated that \emph{any} theory with physical content can be written in generally covariant form. For example, one can \emph{Kretchmannize} Newtonian mechanics by substituting every occurrence of $dt$ in the Newtonian action by $\frac{dt}{d\lambda}d\lambda$, where $\lambda$ is some arbitrary parameter. This gives a reparameterization invariant formulation of Newtonian mechanics often called Parameterized Particle Mechanics (PPM) \cite{lanczos:mechanics, kuchar:canqugr, adm:adm_review}. Though this theory is generally covariant with respect to time, it still contains and absolute temporal structure. GR, as a result, can not be called relational because of its general covariance alone. But if general covariance is not sufficient to make GR relational then what makes it relational? Indeed, is GR relational at all?

The answer to this question, according to Barbour \emph{el al.} \cite{barbourbertotti:mach, barbour_el_al:scale_inv_gravity, barbour_el_al:physical_dof}, is very nearly\footnote{In \cite{barbourbertotti:mach, barbour_el_al:scale_inv_gravity}, GR was found not to be Machian with respect to scale transformations. However, in \cite{barbour_el_al:physical_dof}, it was noticed that GR, in CMC gauge, is Machian with respect to \emph{volume preserving} scale transformations improving the situation considerably.} ``yes''. They were able to derive GR on closed spatial topologies using a technique called \emph{best-matching} which was designed to implement temporal and spatial relationalism. The same techniques applied to models of non-relativistic particles, developed in \cite{barbour:scale_inv_particles, barbourbertotti:mach, barbour:timelessness, gergely:geometry_BB1, gergely:geometry_BB2, anderson:triangleland_old}, lead to temporally and spatially relational theories which suffer from the same Problem of Time encountered in GR. The link between the Problem of Time in GR and in these models is well documented in \cite{barbour:eot, kiefer:qu_gr_book, kuchar:time_int_qu_gr, kuchar:canqugr, adm:adm_review, anderson:rel_part_mech_2, kuchar:prob_of_time, kuchar:lect_notes, barbour:timelessness2} and is discussed at some length in \cite{anderson:triangleland_new_2}. From the perspective of Barbour-Bertotti (BB) theory, the only difference between the relational particle models and GR is the choice of the configuration space\footnote{As well as the \emph{local} square root used in the action.}. They are, thus, useful toy models for studying the Problem of Time in GR\footnote{Of course there are certain features of GR that may be important to the Problem of Time, such as the indefiniteness of the kinetic energy and the fact that it is four dimensional, that are not captured by these toy models.}.

In this paper, we focus specifically on temporal relationalism and its implications to the Problem of Time. We justify this in an \apx{spatial_sym} where we show that adding spatial relationalism by best-matching the spatial symmetries of PPM does not change the main results regarding time. We then propose a new method for implementing temporal relationalism, inspired by best-matching, and apply it to PPM. We prove that this method is equivalent to BB's proposal for non-relativistic particles. Using our proposal, we can study the difference between relational and absolute formulations of PPM. We present two main results: 1) PPM can be made temporally relational by applying our procedure which effectively gauges its time translational symmetry, and 2) the absolute structure can be restored by modifying the variation of the relational fields. In this context, it is the different types of variations of these fields, which are controlled by a constraint we call the Mach constraint, that make the theory either temporally relational or absolute. We thus establish a clear distinction between parameterised theories that have hidden backgrounds and those that don't. By adding a Mach constraint to the time translations, we have been successful at separating the issue of Relationalism (ie, background independence), which is characterized by the Mach constraint, from the issue of reparameterization invariance (ie, general covariance), which is characterized by the Hamiltonian constraint. However, this does not address the deeper issue (considered by \cite{barbour_foster:dirac_thm}) of whether the Hamiltonian constraint should be thought of as a generator of gauge transformations, which arises whenever a theory is reparameterization invariant in the canonical variable.

We arrive at our results using Dirac and path integral quantizations. Though these approaches should, in principle, be equivalent, the path integral approach provides a much more detailed and convincing picture of what is going on.

Given the deep relationship between these toy models and GR, our results indicate that it might be possible to determine an absolute formulation of GR. This would be invaluable for either understanding relationalism in GR or understanding how absolute structures may emerge from quantum gravity. Furthermore, the generalization of our results might shed additional light on the results of \cite{brown:gr_time}.

\subsection{Historical Background}

Barbour and Bertotti have provided a technique \cite{barbourbertotti:mach} called \emph{best-matching} for producing a Machian theory. The key idea is that background structures exist in theories when meaning is attached to the absolute position of the system in a gauge orbit generated by a particular symmetry group. Using a procedure mathematically similar to \emph{gauging} a global symmetry\footnote{A precise relation between best-matching and gauge theory has been proposed in \cite{gryb:ym_bm}.}, the dependence of a theory on a background structure can be eliminated. This procedure is known as the \emph{corrected coordinate method} and is the basis of best-matching.\footnote{The idea of the \emph{corrected coordinate method} itself was introduced after \cite{barbourbertotti:mach} and is developed nicely in \cite{barbour:mach_principle, barbour:scale_inv_particles, barbour_el_al:scale_inv_gravity}.}

Barbour and Bertotti \cite{barbourbertotti:mach} used their technique of best-matching on the simplest non-trivial theory they could write down where the 3-metric is diffeomorphism invariant. This led to the Baierlein-Sharp-Wheeler (BSW) action \cite{bsw:bsw_action} for GR, an approach referred to as \emph{geometrodynamics}. It proves that GR, restricted to $\mathbb{R} \times \Sigma^3$ topologies where $\Sigma^3$ is a compact 3 manifold with no boundary, has no background spatial structure and, as a result, is spatially relational. Temporal relationalism is achieved by implementing Jacobi's principle \cite{lanczos:mechanics}: an action principle that uses the weighted average rate of change of the 3 dimensional configuration space variables multiplied by a suitably defined conformal factor to define a metric on configuration space. As a result, according to Barbour's criterion, Einstein succeeded, though not in the way he originally intended, at finding a spatially and temporally relational theory of gravity.\footnote{GR is \emph{not} relational with respect to local scale transformation of the metric and is, hence, not completely Machian by Barbour's criterion. For formulations of geometrodynamics with and without scale invariance see \cite{barbour_el_al:scale_inv_gravity, barbour_el_al:rel_wo_rel, barbour_el_al:physical_dof, barbourbertotti:mach}.}

The stark asymmetry between how temporal and spatial relationalism is achieved is attributed to a conceptual asymmetry between the nature of time and space. However, in the following paper, we show that for non-relativistic particles, Jacobi's principle is equivalent classically and quantum mechanically to applying the corrected coordinate method to the time translational symmetry of PPM. Thus, Jacobi's principle is seen as originating from the temporal generalization of best-matching. We do not claim that this undermines the conceptual beauty of Jacobi's principle nor the unique role played by time due to the quadratic constraints but, rather, we propose it as a mathematical equivalent for deriving temporal relationalism.

The theory we obtain from applying the corrected coordinate method to PPM we will call time-gauged PPM. As we will show, it is equivalent to Jacobi's principle applied to classical mechanics and is temporally relational as a result. PPM on the other hand, is the Kretschmannized version of Newtonian mechanics and, as such, contains a type of ``hidden'' background structure. Hence, we must make a clear distinction between: 1) manifestly absolute theories (like Newtonian mechanics), 2) Kretschmannized theories with absolute structure (like PPM), and 3) truly relational theories (like time-gauged PPM and GR). It is easy to tell (by definition) theories of type 1) from theories of type 2) and 3). However, it is not always clear how to make a distinction between theories of type 2) and and theories of type 3). For instance, GR, in its formulation due to Arnowitt, Deser, and Misner (ADM), has many formal similarities to PPM. In \cite{adm:adm_review}, ADM point out these similarities suggesting that GR might be a theory of type 2) having a hidden background structure. Solving the problem of time would then involve finding this background structure. This was also a view that has long been supported by Kucha$\check{\text{r}}$ \cite{kuchar:canqugr, kuchar:prob_of_time, kuchar:lect_notes}. In light of Barbour and Bertotti's results, it is not surprising that no attempt to find this background structure has been successful. The results of this paper will suggest that the ADM action might have more in common with time-gauged PPM than standard PPM. In light of this confusion, it would be valuable to have a criterion to distinguishing between type 2 and type 3 theories. We propose such a criterion in the context of time-gauged PPM.

In this work, we study closely what happens when the background structure of PPM is eliminated using the corrected coordinate method. The auxiliary fields, which can be thought of as being related to the flat part of the gauge connection,\footnote{See \cite{gryb:ym_bm} for more details.} are not varied in the standard way but in a way where the endpoints are allowed to vary freely. We will find that it is this free endpoint variation that makes the theory relational. If we use a standard fixed endpoint variation instead, the background structure of the theory is restored. That is, when the corrected coordinate method is applied to the global symmetry of this theory, the difference between the relational theory and the absolute theory is the difference between a free and fixed endpoint variation of the auxiliary fields. This result is tied in to the basic structure of the corrected coordinate method and we expect it to hold generally when any global symmetry is gauged in this way. In particular, we expect it hold also in geometrodynamics where the free endpoint variation of the auxiliary fields is known to lead to GR. This raises the question: what would happen in geometrodynamics if we do a fixed endpoint variation of the auxiliary fields instead? This is still an open question though it may be that the answer is related to the results of \cite{brown:gr_time}.

It is a robust result of best-matching and the corrected coordinate method \cite{barbour:scale_inv_particles} that a relational theory will lead to linear momentum constraints. Classically, as is shown in \cite{gryb:ym_bm}, these constraints guarantee that the Noether current associated with the gauged symmetry is zero. Quantum mechanically it means that the wavefunction is a zero eigenvalue eigenstate of the appropriate generalized momentum. For time-gauged PPM, this means that the wavefunction is an energy eigenstate and does not evolve in time. A similar situation arises in geometrodynamics and leads to the problem of time. However, for a reasonable theory of quantum gravity to agree with everyday experience, one requires time to emerge on isolated subsystems of the universe. How can GR ,which is a relational theory, pick up an absolute structure on isolated subsystems? What would this structure look like? In light of the results mentioned above, we can answer this question in PPM. The answer is that, the kernel for time-gauged PPM should vary the auxiliary fields freely at the endpoints. For local subsystems however, this variation needs to reduce to a variation where the endpoints are approximately fixed. This would introduce an absolute structure. This demonstrates the value of having a definite way to distinguish a theory of type 2) from a theory of type 3).

The confusion surrounding relationalism, which started with Einstein, survives today and these foundational issues still divide the quantum gravity community. This is a particularly important observation when one considers that the difficulties encountered in interpreting the Wheeler-DeWitt equation stem, as just explained, from the temporal relationalism of classical GR. The problem of time, in this sense, can be thought of as the quantum mechanical offspring of the relational versus absolute debate.

\subsection{Outline}

The results of this paper will be presented as follows. In \scn{canqu}, we will develop the full canonical formalism and quantization of time-gauged PPM. On our way, we will study Jacobi-Barbour-Bertotti (JBB) theory (a temporally relational theory of particle mechanics) and PPM as these will be, respectively, the relational and absolute equivalents of time-gauged PPM. We will end the section by considering the fixed endpoint variation and how it reintroduces a background structure. In \scn{pathint}, we repeat the analysis of time-gauged PPM with free and fixed endpoint variations but with phase space path integrals. In this context, there is much intuition to be gained about how time disappears and is reintroduced into the quantum theory. Because of this, the path integral may be a powerful tool for studying how time might emerge for isolated subsystems of the universe.

\section{Canonical Quantization} \label{sec:canqu}

\subsection{Review Of Jacobi-Barbour-Bertotti (JBB) Theory}

We begin with a review of JBB theory and its canonical quantization as a simple example of a temporally relational theory of non-relativistic particle mechanics. The purpose of this review is to lay down the ground work for later calculations. The Hamiltonian constraint of this theory leads to a problem of time just as it does in quantum GR. Though the action for this theory was first written down long ago by Jacobi (for a good treatment see \cite{lanczos:mechanics}) it wasn't until relatively recently that Barbour and Bertotti \cite{barbourbertotti:mach} noticed the significance of the theory in regards to relational time and the problem of time.

\subsubsection{Hamiltonian and Dirac Algebra}

The configuration space of JBB theory is $N\times d$ dimensional space (with $N$ the number of particles and $d$ the number of spatial dimensions) of particle positions $q^i_I$, where lower case indices range from 1 to $d$ and upper case indices label particles and range from 1 to $N$. The dynamics of the system is given by trajectories in configuration space parameterized by the arbitrary parameter $\lambda$. The reparameterization invariant action is:
\equa{\label{eq:jbb_action}
    S_{\text{JBB}} = \int d\lambda\, 2\sqrt{T(\dotq)}\sqrt{E-V(q)},\quad \text{where}\quad T = \sum_I \frac{1}{2} m_I (\dotq_I^i)^2.
}
The canonical momenta are given by
\equa{\label{eq:pjbb}
    p^I_i = \diby{L_\jbb}{\dotq^i_I} = \sqrt{\frac{E-V}{T}} m_I \dotq^j_I \eta_{ij}
}
and are easily seen to obey the quadratic identity
\equa{\label{eq:hamcon}
    \ham = V - E + \sum_I \frac{(p^I_i)^2}{2m_I} =0.
}
This quadratic identity suggests that the momenta should be thought of as directional cosines since the length of the momenta are irrelevant pure gauge degrees of freedom. The canonical Hamiltonian $H_{\text{c}} = \sum_I p^I_i q^i_I - L_{\jbb}$ is zero. Thus, as in any reparameterization invariant theory, the total Hamiltonian is proportional to the constraints. In this case, (\ref{eq:hamcon}) is the only constraint making it first class. This gives us the total Hamiltonian
\equa{\label{eq:hamjbb}
    H_{\text{T}} = N(\lambda) \ham = N(\lambda) \lf(V - E + \sum_I\frac{(p^I_i)^2}{2m_I}\rt).
}

\subsubsection{Equations of Motion}

The fundamental Poisson brackets
\equa{\label{eq:pbjbb}
    \pb{q^i_I}{p^J_j} = \delta^J_I \delta^i_j
}
can now be used to work out Hamilton's first and second equations of motion
\begin{align}
    \dotq^i_I = \pb{q^i_I}{H_{\text{T}}} &= N \frac{p^I_j\eta^{ij}}{m_I}, \quad \text{and} \label{eq:h1jbb}\\
    \dotp_i^I = \pb{p^I_i}{H_{\text{T}}} &= -\diby{V}{q^i_I}\label{eq:h2jbb}.
\end{align}
Combining (\ref{eq:h1jbb}) and (\ref{eq:h2jbb}), we immediately recover Newton's laws for the gauge choice $N=1$. However, because we have the constraint (\ref{eq:hamcon}) in addition to Newton's differential equations, this theory only represents Newtonian mechanics constrained to orbits of constant energy $E$.

\subsubsection{Quantum Theory}\label{sec:quantumJBB}

Following Dirac's procedure, we promote the phase space variables to operators and the fundamental Poisson brackets to commutators. Thus we have (in units where $\hbar = 1$):
\equa{
    \lf[ \opq^i_I, \opp^J_j \rt] = i\, \delta^J_I \delta^i_j.
}
The operators $\opq^i_I$ and $\opp^J_j$ act on the wavefunction $\ket{\Psi}$ which also must obey the operator constraint
\equa{
    \hat\ham \ket{\Psi} = 0
}
or
\equa{
    \lf[ V(\opq) + \sum_I \frac{(\opp^I_i)^2}{2m_I} \rt] \ket{\Psi} = E \ket{\Psi}.
}
This is just the familiar time independent Schr\"{o}dinger equation requiring that the wavefunction of the system must be an energy eigenstate. The Wheeler-DeWitt equation is completely analogous in this sense which is why JBB theory can be used as a toy model for understanding the problem of time in quantum gravity. In this paper we will show that JBB theory is equivalent quantum mechanically to time-gauged PPM. Before gauging the time translational invariance of PPM, we will quickly review it since we will need many of the derived results later.

\subsection{Parameterized Newtonian Mechanics (PPM)}

\subsubsection{Hamiltonian and Dirac Algebra}

The general idea behind PPM is to treat time as a configuration space variable and solve for it dynamically using an action principle (see \cite{lanczos:mechanics, adm:adm_review, henneaux_teit:quant_gauge} for alternative treatments). The configuration space of PPM is then the \emph{extended configuration space} of the particle positions $q^i_I$ and a variable $q^0$ which, once the equations of motion have been solved for, will represent a Newtonian time. It turns out, as we will see, that $q^0$ is also equivalent to Newtonian time off-shell. The ``dynamics'' are given by trajectories in extended configuration space parameterized by the arbitrary parameter $\lambda$. Here $\lambda$ should be thought of as an arbitrary parameter \emph{not} necessarily related to time. The action is:
\equa{\label{eq:actionpnm}
    S_\pnm = \int d\lambda\, \lf[ \frac{T(\dotq)}{\dotq^0} - \dotq^0 V(q) \rt].
}
This action is trivially invariant under translations of the variable $q^0(\lambda) \ra q^0(\lambda) + a$. The momenta conjugate to $q^i_I$ are
\equa{
    p^I_i = \diby{L_\pnm}{\dotq^i_I} = \frac{1}{\dotq^0} m_I \dotq^j_I \eta_{ij}.
}
Comparing this to \eq{pjbb}, we see that the only difference here is that $\dotq^0$ is now playing the role previously played by the quantity $\sqrt{\frac{E-V}{T}}$. Indeed, in JBB theory $\sqrt{\frac{E-V}{T}}$ is typically identified with the $\lambda$ derivative of the \emph{ephemeris time} $\tau_{\text{EPH}}$ \cite{barbourbertotti:mach}. However, in PPM, $\dotq^0$ is a configuration space variable while the ephemeris time is \emph{not} in JBB theory. This will have important ramifications for the role played by the different kinds of time variables in the quantum theory.

In extended configuration space there is a momentum conjugate to $q^0$ leading to an extended phase space. This momentum is
\equa{
    p_0 = \diby{L_\pnm}{\dotq^0} = - \lf( \frac{T}{(\dotq^0)^2} + V \rt).
}
The momenta obey the identity
\equa{\label{eq:hamconpnm}
    \ham = p_0 + V + \sum_I \frac{(p^I_i)^2}{2m_I} = 0.
}
Again, the canonical Hamiltonian $H_{\text{c}} = \dotq^0 p_0 + \sum_I p^I_i q^i_I - L_{\pnm}$ is zero which means that the total Hamiltonian $H_{\text{T}}$ is proportional to the lone first class constraint (\ref{eq:hamconpnm})
\equa{
    H_{\text{T}} = N(\lambda) \ham = p_0 + V + \sum_I \frac{(p^I_i)^2}{2m_I}.
}
The only (but crucial) difference between this Hamiltonian and the one in (\ref{eq:hamjbb}) for JBB theory is that $p_0$ is a phase space variable instead of a constant $-E$. This will lead to a completely different quantum theory since phase space variables are promoted to operators but constants are not. Because of this, quantum PPM will have an absolute time and be free of a problem of time.

\subsubsection{Equations of Motion}

In addition to the fundamental Poisson brackets $\pb{q^i_I}{p^J_j} = \delta^J_I \delta^i_j$, which are identical to the Poisson brackets (\ref{eq:pbjbb}) of JBB theory, we also have the Poisson brackets
\equa{
    \pb{q^0}{p_0} = 1.
}
Using these, we find that the equations of motion for the $\dotq$'s and the $\dotp$'s are completely identical to (\ref{eq:h1jbb}) and (\ref{eq:h2jbb}) of JBB theory. However, in PPM we have additionally
\begin{align}
    \dotq^0 = \pb{q^0}{H_{\text{T}}} &= N(\lambda), \quad \text{and} \label{eq:h1pnm} \\
    \dotp_0 = \pb{p_0}{H_{\text{T}}} &= 0. \label{eq:h2pnm}
\end{align}
\eq{h2pnm} expresses conservation of energy while \eq{h1pnm} tells us that picking a function $N(\lambda)$ is tantamount to picking a time gauge. In light of this, the $N=1$ gauge can be thought of as the \emph{Newtonian gauge} as it is the gauge where Newton's laws are manifestly valid.

\subsubsection{Quantum Theory}

Promoting all the phase space variables to operators and Poisson brackets to commutators gives the operator algebra
\equa{
    \lf[ \opq^i_I, \opp^J_j \rt] = i\, \delta^J_I \delta^i_j, \quad \text{and} \quad \lf[ \opq^0, \opp_0 \rt] = i.
}
The operators act on the ket $\ket{\Psi}$ which obeys the operator constraint
\equa{
    \hat\ham \ket{\Psi} = \lf[ V(\opq) + \opp_0 + \sum_I \frac{(\opp^I_i)^2}{2m_I} \rt] \ket{\Psi} = 0.
}
Associating $\opq^0$ with a time operator, this is just the familiar time dependent Schr\"{o}dinger equation. Thus, the quantization of PPM leads to standard quantum theory for non-relativistic particles with no problem of time. The main reason for this is that PPM contains a ``hidden'' background structure that it inherited from standard Newtonian mechanics. It is possible to remove this background structure however by applying a procedure that will be treated in the next section.

\subsection{The Corrected Coordinate Method and PPM}

The corrected coordinate method is a technique developed for best-matching by Barbour and collaborators \cite{barbour:scale_inv_particles, barbour_el_al:scale_inv_gravity} to remove a theory's dependence on some non-physical absolute structure. One can think of it as a way of gauging a global symmetry associated to the absolute structure in question. The link between the corrected coordinate method and standard gauge theory has been explored in \cite{gryb:ym_bm}. For our purposes, the main result that we will need is that the corrected coordinate method will eliminate the dependence of our theory on the non-physical part of the gauge fields associated with a global symmetry. In this case, the global symmetry we are interested in is the time translational invariance of PPM. Gauging this symmetry will reduce the theory to JBB theory.

\subsubsection{The Method}

If the configuration space coordinates contain a physical symmetry under the transformation $q^i_I \ra G(\omega^\alpha)^i_j q^j_I$, where $G$ is a matrix representation of the group symmetry and $\omega^\alpha$ are group parameters, then the corrected coordinate method involves replacing the coordinates $q^i_I$ with the \emph{corrected coordinates} $\bar{q}^i_I = G(\omega'^\alpha)^i_j q^j_I$. Here, $\omega'^\alpha$ are dynamical fields called \emph{auxiliary fields}. By construction, they do not contribute to the true degrees of freedom of the system but shift the physical degrees of freedom along gauge orbits. The dynamics of the auxiliary fields is determined by a \emph{free} endpoint variation of the action since the value of the auxiliary fields on the boundary is just as arbitrary as it is in the bulk. In a free endpoint variation we must require the additional constraint
\equa{
    \diby{L}{\dot{\omega}'^\alpha} = 0
}
on top of the usual Euler-Lagrange equations. In the Hamiltonian formulation, this is equivalent to imposing the additional free endpoint condition
\equa{\label{eq:machconstraint}
    \pi^\alpha = 0,
}
where $\pi^\alpha$ are the momenta conjugate to $\omega'_\alpha$, on top of the usual Hamilton equations of motion. I will call (\ref{eq:machconstraint}) the \emph{Mach constraint}. We will see how the role of the Mach constraint is to eliminate the background structure of a theory leaving a true relational theory behind. In best-matching, the Mach constraint leads to linear momentum constraints which can be solved for to eliminate the auxiliary fields and obtain a theory in terms of the physical degrees of freedom of the system. Here we will apply the corrected coordinate method to gauge the time translational invariance of PPM in an attempt to introduce a relational time.

\subsubsection{Hamiltonian and Dirac Algebra} \label{sec:diracpnm}

We introduce the corrected coordinates
\equa{
    \bar{q}^0(\lambda) = G(a(\lambda))q^0 = q^0(\lambda) + a(\lambda)
}
in order to apply the corrected coordinate method to the time translational invariance of PPM. This further extends the configuration space to include the auxiliary field $a(\lambda)$. Next, we substitute $q^0 \ra \bar{q}^0$ in the action of (\ref{eq:actionpnm}) giving
\equa{\label{eq:action}
    S = \int d\lambda \lf[ \frac{T}{\dotq^0 + \dota} - (\dotq^0 + \dota)V \rt].
}
The canonical momenta are given by
\begin{align}
    p^I_i &= \diby{L}{\dotq^i_I} = \frac{1}{\dotq^0 + \dota} m_I \dotq^j_I \eta_{ij} \\
    p_0 &= \diby{L}{\dotq^0} = - \lf( \frac{T}{(\dotq^0+ \dota)^2} + V \rt). \label{eq:ppm_p0}
\end{align}
Notice that these are the same as the momenta for PPM with the replacement $\dotq^0 \ra \dotq^0 + \dota$. However, we now have an additional momentum conjugate to $a$. It is given by
\equa{
    \pi = \diby{L}{\dota} = - \lf( \frac{T}{(\dotq^0+ \dota)^2} + V \rt).\label{eq:ppm_pi}
}
This introduces the new momentum constraint
\equa{
    \lin \equiv p_0 - \pi = 0.
}
We will call this the \emph{linear momentum constraint} in analogy with the diffeomorphism constraint of quantum General Relativity. Because we are doing a free endpoint variation of the $a$'s we must also impose the Mach constraint
\equa{
    \pi = 0.
}
Finally, there is also a Hamiltonian constraint
\equa{
    \ham = p_0 + V + \sum_I \frac{(p^I_i)^2}{2m_I} = 0
}
which is identical to the Hamiltonian constraint of PPM. Collecting our results, we have three constraints $\lin = \pi = \ham =0$ two linear and one quadratic in the momenta. The constraints that are linear in the momenta are very straightforward and one can solve for them and integrate them out. However, it is very enlightening to keep them around so that we can see how they integrate out explicitly. This will reveal the full structure of the theory and will give us conceptual clues to how time disappears. Also, it will provide a good model for what we should expect to happen in more complicated theories like best-matched particle mechanics or general relativity where the constraints are not as easy to solve for.

The canonical Hamiltonian $H_{\text{c}}$, can once again be shown to be zero. This leaves a total Hamiltonian which is a linear combination of the constraints
\begin{align}
    H_{\text{T}} &= N(\lambda) \ham + L(\lambda) \lin + M(\lambda) \pi \notag \\
                 &= N \lf[ p_0 + V + \sum_I \frac{(p^I_i)^2}{2m_I} \rt] + L [p_0 - \pi] + M\pi.
\end{align}

We can work out the Dirac algebra by using the fundamental Poisson brackets
\begin{equation}
\begin{array}{ccc}
  \pb{q^i_I}{p^J_j} = \delta^J_I \delta^i_j, & \pb{q^0}{p_0} = 1, \text{and} & \pb{a}{\pi} = 1.
\end{array}
\end{equation}
Since the constraints commute with themselves, we are left to work out the Poisson brackets between different constraints. It is a very short calculation to show that
\begin{align}
    \pb{\ham}{\lin} &=0 \\
    \pb{\ham}{\pi} &=0, \quad\text{and} \\
    \pb{\pi}{\lin} &=0.
\end{align}
Thus, all the constraints are first class.

\subsubsection{Gauge Transformations}

Because $\lin = 0$ and $\pi = 0$ are ordinary linear momentum constraints they will generate gauge transformations. For discussions of the action of the quadratic constraint of parameterized theories, such as $\ham = 0$, see \cite{barbour_foster:dirac_thm}. Here we will only be concerned with the gauge transformations generated by ordinary constraints.

We can see the types of gauge transformations generated by the linear constraints by computing the infinitesimal change generated by each of the constraints on the configuration space variables. For $\lin$ this is
\begin{align}
    \delta_\lin q^i_I & = \epsilon(\lambda) \pb{q^i_I}{q^0 - \pi} =0  \\
    \delta_\lin q^0 & = \epsilon(\lambda) \pb{q^0}{q^0 - \pi} = \epsilon(\lambda) \\
    \delta_\lin a & = \epsilon(\lambda) \pb{a}{q^0 - \pi} =-\epsilon(\lambda).
\end{align}
Hence, $\lin$ generates the transformation $q^0 \ra q^0 + \epsilon$ and $a \ra a - \epsilon$ which is clearly an invariance of the action (\ref{eq:action}). Because of its triviality, it has been called the \emph{banal invariance} by Barbour \cite{barbour:scale_inv_particles}. However, by keeping all the constraints we can see that there is yet another even more trivial invariance of the action which I will call the \emph{Machian invariance}. It is generated by the Mach constraint $\pi$
\begin{align}
    \delta_\pi q^i_I & = \epsilon(\lambda) \pb{q^i_I}{\pi} =0  \\
    \delta_\pi q^0 & = \epsilon(\lambda) \pb{q^0}{\pi} = 0 \\
    \delta_\pi a & = \epsilon(\lambda) \pb{a}{\pi} =\epsilon(\lambda).
\end{align}
which is given by the transformation $a \ra a + \epsilon$. The reason for the name is that this invariance guarantees that the physical theory is invariant under infinitesimal translations of the $a$'s. This is another way of saying that the $a$'s are arbitrary and therefore cannot affect the physical content of the theory. It is a requirement of a relational theory.

\subsubsection{Equations of Motion}\label{sec:eom}

Using the fundamental Poisson brackets we can easily work out Hamilton's equations of motion for the system. The equations of motion for the spatial variables are unchanged from PPM. They are given by \eq{h1jbb} and \eq{h2jbb}. There are differences however in the equations of motion of the time variables and the auxiliary fields. For the time variables we have
\begin{align}
    \dotq^0 &= \pb{q^0}{H_{\text{T}}} = N + L \label{eq:h1}\\
    \dotp_0 &= \pb{p_0}{H_{\text{T}}} = 0 \label{eq:h2}
\end{align}

We still have conservation of energy from \eq{h2} and an arbitrary $\dotq^0$ from \eq{h1} but now the Lagrange multiplier of \eq{h1} is different, in general, from that of \eq{h1pnm}. The equations of motion of the auxiliary fields are
\begin{align}
    \dota &= \pb{a}{H_{\text{T}}} = M - L \label{eq:h1a}\\
    \dot{\pi} &= \pb{\pi}{H_{\text{T}}} = 0 \label{eq:h2a}.
\end{align}
\eq{h1a} says that $\dota$ is arbitrary. We will use these equations of motion to determine expressions for the gauge fixing functions in the path integral quantization.

In JBB theory, the only variables we have are the $q^i_I$ and the $p_i^I$. They obey the same equations of motion as the $q^i_I$ and the $p_i^I$ of this theory. Furthermore, classically the constraints on the momenta $\lin = \pi = \ham =0$ can be reduced to
\equa{
    \lf(V + \sum_I\frac{(p^I_i)^2}{2m_I}\rt) =0
}
which is identical to the Hamiltonian constraint of \eq{hamcon} with $E=0$. However, the energy can easily be absorbed into the constant part of the potential $V_0$ restoring the equivalence of the constraints. Therefore, as far as the spatial variables are concerned, the two theories are classically equivalent. The only difference is that here we have additional auxiliary fields $a$ which do not affect the physical content of the theory. Their role can be thought of as a way of eliminating $p_0$ from the Hamiltonian constraint.

\subsubsection{Quantum Theory}\label{sec:qupnm}

Promoting the phase space variables to operators and the Poisson brackets to commutators we have
\equa{
\begin{array}{ccc}
   \lf[ \opq^i_I, \opp^J_j \rt] = i\, \delta^J_I \delta^i_j, & \lf[ \opq^0, \opp_0 \rt] = i, \quad\text{and} & \lf[ a, \pi \rt] = i.
\end{array}
}
The operators act on the wavefunction $\ket{\Psi}$ which must obey the constraints
\equa{
  \hat\ham \ket{\Psi} = \hat\lin \ket{\Psi} = \hat\pi \ket{\Psi} = 0.
}
That is, the wavefunction must simultaneously satisfy
\begin{align}
    \lf[ V(\opq) + \sum_I \frac{(\opp^I_i)^2}{2m_I} \rt] \ket{\Psi} &= -\opp_0 \ket{\Psi},  \label{eq:hamconspnmbm}\\
    \opp_0 \ket{\Psi} &= \hat\pi \ket{\Psi}, \label{eq:lincons}\qand \\
    \hat\pi \ket{\Psi} &= 0.
\end{align}
Combining these leads immediately to the time independent Schr\"{o}dinger equation
\equa{
    \lf[ V(\opq) + \sum_I \frac{(\opp^I_i)^2}{2m_I} \rt] \ket{\Psi} = 0
}
with energy $E$ equal to zero. The fact that the energy here appears to be zero is slightly deceiving since the potential $V$ could have a constant term $V_0$ which could be interpreted as the negative of the energy. Thus, this is not necessarily a theory with zero energy. Nevertheless, we have still made the connection quantum mechanically with the quantum version of JBB theory outlined in \scn{quantumJBB}.

\subsubsection{Fixed Endpoint Variation}

We will end this section by solving the interesting problem of using a \emph{fixed} endpoint variation rather than a free endpoint variation when implementing the corrected coordinate method on the time translational invariance of PPM. We will attempt to mirror the discussion of the free endpoint variation of the earlier part of this section.

The phase space and action used in the fixed endpoint variation are identical to those of the free endpoint variation. As a result, the expressions for $p^I_i$, $p_0$, and $\pi$ are identical to the expressions given in \scn{diracpnm}. Furthermore, because the definitions of the momenta are unchanged, they will obey the same identities
\begin{align}
    \ham &= p_0 + V + \sum_I \frac{(p^I_i)^2}{2m_I} = 0, \qand \\
    \lin &= p_0 -\pi = 0.
\end{align}
However, because we would like to fix the variation at the endpoints we need to drop the Mach constraint. This will lead to the standard canonical analysis of the system. The total Hamiltonian $H_{\text{T}}$ of the system is
\equa{
    H_{\text{T}} = N\ham + L\lin.
}
Having one less constraint in the system means we have only one non-trivial Poisson bracket, $\pb{\ham}{\lin}$, to work out. It is easily seen to be zero. Thus, our two constraints are first class.

Being first class, $\lin$ will generate gauge transformations. Because the fundamental Poisson brackets are unchanged, $\lin$ will still generate the banal transformations
\begin{align}
    q^0 &\ra q^0 + \epsilon, \qand \\
    a &\ra a - \epsilon.
\end{align}
and the Hamiltonian constraint $\ham$ will behave as it did before. However, the absence of a Mach constraint implies that we no longer have a Machian invariance. Thus, if one takes the terminology seriously, we no longer will have a Machian theory. We will now see why one should take this terminology seriously.

If we work out the equations of motion, we find
\begin{align}
    \dotq^i_I &= \pb{q^i_I}{H_{\text{T}}} = N \frac{p^I_j \eta^{ij}}{m_I}, & \dotp_i^I &= \pb{p_i^I}{H_{\text{T}}} = -\diby{V}{q^i_I}, \\
    \dotq^0 &= \pb{q^0}{H_{\text{T}}} = N + L, & \dotp_0 &= \pb{p_0}{H_{\text{T}}} = 0, \\
    \dota &= \pb{a}{H_{\text{T}}} = - L,\qand & \dot{\pi} &= \pb{\pi}{H_{\text{T}}} = 0.
\end{align}
The gauge freedom provided by the surviving banal invariance allows us to pick a gauge where $L = 0$. In this gauge, it is manifest that this system is classically equivalent to PPM since the equations of motion are identical and the $a$'s are irrelevant variables due to the banal invariance. Since PPM is really just Newton's theory (this can easily be seen by setting $N=1$) it is clear that we now have a theory containing a background absolute structure for time. This is our first indication that the presence of the Mach constraint, or alternatively a free endpoint variation on the auxiliary fields, is the definition of a Machian theory. Its presence provides a clear criterion to distinguish between a Kretchmanized theory, or generally covariant theory, and a true relational theory. We will see now and in the next sections how this criterion is also good at the quantum level.

Let us briefly examine the canonical quantization of the theory first, showing that fixing the endpoints restores an absolute structure to the quantum theory, then provide a much deeper analysis using path integrals that will lead to the same results but will provide us with more intuition. The quantum mechanical operators and commutators are identical to those of \scn{qupnm}. However, now the wavefunction $\ket{\Psi}$ need only obey the two constraints
\begin{align}
    \lf[ V(\opq) + \sum_I \frac{(\opp^I_i)^2}{2m_I} \rt] \ket{\Psi} &= -\opp_0 \ket{\Psi},\label{eq:hamfixed}\qand \\
    \opp_0 \ket{\Psi} &= \hat\pi \ket{\Psi}. \label{eq:linfixed}\\
\end{align}
\eq{linfixed} says that the operators $\opp_0$ and $\hat\pi$ are interchangeable. The banal invariance gives us the freedom to make $\hat{q}^0$ and $\hat a$ interchangeable making the auxiliary fields somewhat redundant. We are left with \eq{hamfixed} which is just the time \emph{dependent} Schr\"{o}dinger equation. This makes it clear, that, in the quantum theory, the time dependence has also been restored. We will see that this is true in much more detail when we explore the path integral quantization of this system.

%In this case, we need only drop the Mach constraint. What we are left with is the Hamiltonian constraint (\ref{eq:hamconspnmbm}) which is identical to the Hamiltonian constraint (\ref{eq:hamconpnm}) of PPM and the linear momentum constraint (\ref{eq:lincons}). The linear constraint does not affect the local physics since the $a$'s are only fixed on the boundary. Thus, we left only with the Hamiltonian constraint (\ref{eq:hamconspnmbm}) which is just the time \emph{dependent} Schr\"{o}dinger equation. We see that the theory is now equivalent to standard time dependent quantum mechanics plus some arbitrary gauge fields. The issue of the auxiliary fields on the boundary will be addressed again in more detail when we treat the path integral quantization.

\subsection{Results and Discussions}

We have seen that time-gauged PPM is classically and quantum mechanically equivalent to JBB theory. The role of the auxiliary fields and the Mach constraint is, classically, to remove $p_0$ from the Hamiltonian constraint of PPM. Quantum mechanically, this leads to the time \emph{independent} \schro equation rather then its time \emph{dependent} counterpart. We have worked out the Dirac algebra and found the first class constraints. In a Machian, theory there is a Machian invariance associated with the Mach constraint. If this constraint is lifted, the background structure is restored leaving a theory that is no longer Machian despite still being reparameterization invariant. This suggests to us a criterion for distinguishing true relational theories from reparameterization invariant theories. We will now study in more detail the process through which time disappears from the quantum theory during the gauging process by considering the path integral of time-gauged PPM.

\section{Path Integral Quantization}\label{sec:pathint}

To construct the path integral for time-gauged PPM, which, as we have seen, is a gauge theory, we would like to start from the phase space path integral and apply the procedure of Faddeev and Popov for determining the measure to correctly divide out by the gauge volume \cite{faddeev:fp}. Because of the reparameterization invariance of the action there are some subtleties in how to deal with the gauge invariance but many of these details have been worked out in \cite{sg:emer_time_PI}. We will compare the kernel we get to that obtained from JBB theory and will find it to be identical. Furthermore, we will analyze what happens in the path integral language when the auxiliary fields are fixed on the endpoints. Doing this will introduce a new absolute structure into the theory leading to a kernel that is identical to that of PPM. This will provide further evidence that one can switch between a relational theory and a theory with a background structure by switching between free and fixed endpoints respectively. The path integral does not add new conclusions to the results already derived but it provides much deeper insight into what is going on. Furthermore, it provides a powerful mathematical tool for studying the quantum mechanics of the system.

\subsection{Free Endpoint Time-Gauged PPM}\label{sec:freepnm}

In the phase space path integral, we must integrate over the entire phase space which, in this case, constitutes the spatial coordinates $q^i_I$ for each particle, the time coordinate $q^0$, the auxiliary field $a$, and the momenta conjugate to each of these fields. Because this is a gauge theory, the momenta satisfy constraints. In this case, we have the three constraints derived in \scn{diracpnm}\footnote{An alternative starting point for the free endpoint variation of the auxiliary fields is to remove the Mach constraint but integrate over all possible endpoints for $a$. This method leads to the same result as the method outlined in this section but does not illustrate the role of the Mach constraint which is necessary in the canonical approach. See \apx{A} for the details of the derivation without the Mach constraint.}:
\begin{align}
	\ham &= p_0 + V + \sum_I \frac{(p^I_i)^2}{2m_I} = 0, \\
	\lin &= p_0 - \pi = 0, \qand \\
	\pi  &= 0.
\end{align}
To satisfy these constraints we add them to the Hamiltonian with Lagrange multipliers $N$, $L$, and $M$. These constraints restrict the system to lie on a hypersurface of three less dimensions on momentum space. To resolve the ambiguity caused by the many to one map between the $\dotq$'s and the $p$'s, we must further impose gauge fixing constraints for each of the first class constraints. These constraints must uniquely specify $\dotq$ for a given $p$ and will constrain the system to lie on a hypersurface of 6 less dimensions on phase space (that is, it will be a phase space with 3 less configuration variables). There will be a map from this hypersurface to the true physical phase space of the system that will induce a measure on the hypersurface. This measure must be invariant under canonical transformations and redefinitions of the constraints. The measure satisfying these requirements, proposed by Faddeev and Popov, is the determinant of the Poisson bracket between the first class constraints and their gauge fixing conditions.

In \scn{gaugefix} to \scn{boundarycond} we will set up the technical details necessary to perform the integrations which reduce the kernel to JBB form. Once these are out of the way, in \scn{evalfree} we will focus on how to carry out the integrations paying special attention to the physical processes involved and the disappearance of time.

\subsubsection{Gauge Fixing Conditions}\label{sec:gaugefix}

Inspired by Hamilton's first equations of motion, derived in \scn{eom}, we propose the following natural gauge fixing conditions
\begin{align}
	\calg &= \sum_I \frac{m_I \dotq^i_I p^I_i}{p_I^2} - f = 0,\\
	\calf &= \dotq^0 - g = 0, \qand \\
	\calp &= \dota - h = 0,
\end{align}
where $f$, $g$, and $h$ are \emph{arbitrary} functions on phase space. These constraints will specify a unique $\dotq$ for a given $p$. The arbitrariness of $f$, $g$, and $h$, represents the gauge freedom inherent in the system. Thus, the action in terms of the physical degrees of freedom should not depend on the choice of these functions. To impose these constraints, we add them to the total Hamiltonian of the system with the Lagrange multipliers $\Eps$, $R$, and $S$ respectively.

\subsubsection{Kernel}

The phase space path integral is now an integration over all possible trajectories in phase space for all possible values of $L$, $M$, $N$, $\Eps$, $R$, and $S$ with a measure given by the Faddeev-Popov determinant. The canonical Lagrangian is
\begin{align}
    L &= \sum_I \dotq^i_I p^I_i + \dotq^0 p_0 + \dota \pi - H_{\text{T}} \notag \\
      &= \sum_I \dotq^i_I p^I_i + \dotq^0 p_0 + \dota \pi - N\ham - L\lin - M\pi - \Eps\calg - R\calf -S\calp.
\end{align}
where the $\lambda$ derivatives are taken using the definition $\dotq^i_I(\lambda) = \frac{q^i_I(\lambda + d\lambda) - q^i_I(\lambda)}{d\lambda}$ (and similar formulas for $\dotq^0$ and $\dota$) so that $L$ is a function of phase space. The kernel, $K$, is a function of the differences $Q^i_I \equiv q^i_I(\lambda_{\text{final}}) - q^i_I(\lambda_{\text{initial}})$ and $\tau \equiv q^0(\lambda_{\text{final}}) - q^0(\lambda_{\text{initial}})$ which are the endpoints of the trajectories in extended configuration space. Because we are performing a free endpoint variation for the auxiliary fields, the Mach constraint should eliminate the dependence of the kernel on any boundary data that could be specified for them. We will see explicitly how this happens later. Collecting all this information, we have
\begin{equation}\label{eq:kernelmaster}
    K(Q^i_I, \tau) = \int \cald q^i_I\, \cald p^I_i\, \cald q^0\, \cald p_0\, \cald a\, \cald \pi\, \cald N\, \cald M\, \cald L\, \cald \Eps\, \cald R\, \cald S\, \abs{\pb{\chi_a}{\chi_b}} e^{i \int d\lambda\, L}.
\end{equation}
In the above expression, we have collected all the constraints in the variable $\chi_a$ where $a$ and $b$ range over all the constraints.

\subsubsection{Faddeev-Popov Determinant and Partial Gauge Fixing}

To make the Faddeev-Popov determinant easier to compute and deal with we would like to do a partial gauge fixing. In general, the functions $f$, $g$, and $h$ can be arbitrary functions of phase space but this would make the Faddeev-Popov determinant a horrible mess. Instead, we would like to require that the gauge fixing function $f$ be a function only of the spatial coordinates $q^i_I$ and their conjugate momenta $p_i^I$ while the functions $g$ and $h$ be functions of $\lambda$ only. This allows us to simplify the determinant to the expression
\equa{
    \abs{\pb{\chi_a}{\chi_b}} = \abs{\pb{\ham}{\calg}} = \abs{\pb{V + \sum_I \frac{p_I^2}{2m_I}}{\sum_J m_J \frac{\dotq^i_J p^J_i}{p_I^2}-f}}.
}
This is still slightly non-trivial but we will only need to write it down formally as we are not interested in solving the full path integral but, rather, just enough of it to be able to compare it to the kernel of JBB theory.

\subsubsection{Boundary Conditions}\label{sec:boundarycond}

To evaluate the path integral we break up each path integration into a discrete product of integrals. Slicing each trajectory into $N$ pieces plus 2 endpoints, we replace the functional dependence on $\lambda$ by a discrete Greek index that can range from 0 to $N$. Integrations over the Lagrange multipliers are normalized to give $\delta$-functions.

For all the configuration space variables $q^i_{I,\alpha}$, $q^0_\alpha$, and $a_\alpha$, the endpoints are fixed by the boundary conditions. This would normally mean that we have $N-1$ integrations instead of the $N$ integrations necessary for the remaining variables (which include the momenta and their Lagrange multipliers). However, for each of the first class constraints we have gauge fixing functions which eliminate one degree of freedom in configuration space. Thus, for these degrees of freedom, we need a different prescription for fixing the boundary conditions. In this case, instead of just dropping an integration and fixing the endpoints to satisfy the boundary conditions, we keep the extra integration then pick gauge fixing functions that guarantee the boundary conditions are satisfied. Because the gauge fixing functions uniquely specify the configuration space variable in question, it is possible to chose them in such a way that the boundary conditions are satisfied. This will amount to keeping an extra integration over the configuration space variable in question then imposing a constraint on the gauge fixing functions.

In our case, we have 3 gauge fixing functions which correspond to the specification of 3 configuration space degrees of freedom. We will now implement the method described above for each of these 3 variables.

\begin{enumerate}
	\item The constraint $\calf$ tells us that the gauge fixing function $g$ determines the configuration space variable $q^0$. The boundary condition for this variable is simply
	\equa{\label{eq:q0const}
		\sum_{\alpha = 0}^{N-1} \dotq^0_\alpha \Delta\lambda_\alpha = q^0(\lambda_{\text{final}}) - q^0(\lambda_{\text{initial}}) \equiv \tau_q.
	}
	where we use the definition $\dotq^0_\alpha \equiv \frac{q^0_{\alpha+1}-q^0_{\alpha}}{\Delta\lambda_\alpha}$. It is clear from this that choosing $g_\alpha$ such that
	\equa{\label{eq:g0const}
		\sum_{\alpha = 0}^{N-1} g_\alpha \Delta\lambda_\alpha = \tau_q
	}
	will guarantee that (\ref{eq:q0const}) is satisfied. The calculations are greatly simplified by noting that $q^0$ is cyclic (that is, it enters the action only through its derivative $\dotq^0$). This suggests the change of variables with unit determinant $q^0_\alpha \ra q^0_{\alpha+1}-q^0_{\alpha} = \dotq^0_\alpha \Delta\lambda_\alpha$. Because we are allowing all the variables to vary freely, all integrations are still from $-\infty$ to $\infty$.
	\item The constraint $\calp$ tells us that the gauge fixing function $h$ determines the configuration space variable $a$. Imposing boundary conditions on $a$ may seem odd given the fact that we would like to vary $a$ freely at the endpoints. However, the power of the Mach constraint $\pi=0$ is that, even if we try to impose boundary conditions on $a$, the extra gauge freedom introduced by the arbitrary function $h$ should be enough to eliminate the theory's dependence on them. Thus, to test that the Mach constraint is doing its job we will try to impose the conditions
	\equa{\label{eq:aconst}
		\sum_{\alpha = 0}^{N-1} \dota_\alpha \Delta\lambda_\alpha = a(\lambda_{\text{final}}) - a(\lambda_{\text{initial}}) \equiv \tau_a,
	}
	where we use the definition $\dota_\alpha \equiv \frac{a_{\alpha+1}-a_{\alpha}}{\Delta\lambda_\alpha}$, and see if $\tau_a$ disappears from the final theory. It is clear from the above that choosing $h_\alpha$ such that
	\equa{\label{eq:h0const}
		\sum_{\alpha = 0}^{N-1} h_\alpha \Delta\lambda_\alpha = \tau_a
	}
	will guarantee that (\ref{eq:aconst}) is satisfied. Again, we will use the variable substitution $a_\alpha \ra a_{\alpha+1} - a_\alpha= \dota_\alpha \Delta\lambda_\alpha$ to simplify the integrations.
    \item The most difficult boundary condition to implement using gauge fixing functions is the one on the $q^i_I$. This condition will not be needed when comparing the kernel of best-matched PPM to that of JBB theory as we won't need to integrate over all the degrees of freedom in order to compare the two theories. For this reason, we will delay this discussion referring the reader to \cite{sg:emer_time_PI} to see how this can be done.
\end{enumerate}

\subsubsection{Evaluating the Path Integral}\label{sec:evalfree}

Our goal is to show that the kernel (\ref{eq:kernelmaster}) is equivalent to the kernel of JBB theory. In order to compare the two theories, we must integrate (\ref{eq:kernelmaster}) over $\cald q^0$, $\cald p_0$, $\cald a$, $\cald \pi$, $\cald M$, $\cald L$, $\cald R$, and $\cald S$. The order in which these integrations are done is irrelevant but we will sketch an option that will attempt to highlight the physical processes occurring in the path integrations.

Solving first for $g_0$ using (\ref{eq:g0const}) and $h_0$ using (\ref{eq:h0const}) we will rewrite the kernel of (\ref{eq:kernelmaster}) for completeness specifying the measure exactly using the discretization:
\begin{multline}\label{eq:longkernel}
    K(Q^i_I, \tau_q,\tau_a) = \int_{-\infty}^\infty \frac{d^d p_i^{I,0}}{(2\pi)^d} \frac{dN_0}{2\pi} d \dotq^0_{0} \frac{dR^0}{2\pi} \frac{d p_0^{0}}{2\pi} \frac{dL_0}{2\pi} d \dota_{0} \frac{dS^0}{2\pi} \frac{d \pi^{0}}{2\pi} \frac{dM_0}{2\pi} \times \exp \lf\{ i \lf( S_0\tau_a + R_0\tau_q \rt) \rt\} \\
    \times \exp \lf\{ i \Delta\lambda_0 \lf[ \sum_I \dotq^i_{I,0} p^{I,0}_i + \dotq^0_0 p_0^0 + \dota_0 \pi^0 - N_0\lf( p_0^0 + V^0 + \sum_I \frac{p_{I,0}^2}{2m_I} \rt) - L_0(p_0^0 - \pi^0) \rt.\rt. \\
    \lf. \lf. - R^0 \dotq^0_0 - M_0\pi^0 - S^0\dota_0 \rt] \rt\} \prod_{\alpha = 1}^{N-1} d^d q^i_{I,\alpha} \frac{d\Eps^\alpha}{2\pi} \frac{d^d p_i^{I,\alpha}}{(2\pi)^d} \frac{dN_\alpha}{2\pi} d \dotq^0_{\alpha} \frac{dR^\alpha}{2\pi} \frac{d p_0^{\alpha}}{2\pi} \frac{dL_\alpha}{2\pi} d \dota_{\alpha} \frac{dS^\alpha}{2\pi} \frac{d \pi^{\alpha}}{2\pi} \frac{dM_\alpha}{2\pi} \\
    \times \exp \lf\{ i \Delta\lambda_\alpha \lf[ \sum_I \dotq^i_{I,\alpha} p^{I,\alpha}_i + \dotq^0_\alpha p_0^\alpha + \dota_\alpha \pi^\alpha - N_\alpha\lf( p_0^\alpha + V^\alpha + \sum_I \frac{p_{I,\alpha}^2}{2m_I} \rt) - \Eps^\alpha \lf( \sum_I \frac{m_I \dotq^i_{I,\alpha} p^{I,\alpha}_i}{p_{I,\alpha}^2} - f_\alpha \rt) \rt.\rt. \\
    \lf. \lf. - L_\alpha(p_0^\alpha - \pi^\alpha) - R^\alpha (\dotq^0_\alpha - g_\alpha) -R^0 g_\alpha - M_\alpha\pi^\alpha - S^\alpha (\dota_\alpha - h_\alpha) - S^0 h_\alpha \rt] \rt\}\times \abs{\pb{\ham}{\calg}}.
\end{multline}

Here we have made the coordinate transformations $q^0_\alpha \ra q^0_{\alpha+1}-q^0_{\alpha} = \dotq^0_\alpha \Delta\lambda_\alpha$ and $a_\alpha \ra a_{\alpha+1} - a_\alpha= \dota_\alpha \Delta\lambda_\alpha$ discussed in \scn{boundarycond}. Though this is a long expression it is important to be explicit about the differences between the $\alpha = 0$ terms and the $\alpha = 1,\hdots,N-1$ terms as these differences contain all the information about the boundary conditions which, in turn, contain all the information about time. In the above expression, we insert the Faddeev-Popov determinant only formally since the partial gauge fixing we have used will make it independent of the integrations we are going to perform in order to bring the kernel into JBB form.

We will perform the integrations in two separate steps. The first step will eliminate the kernel's dependence on $\tau_a$ and the arbitrary functions $h_\alpha$. This must be the case since the Mach constraint $\pi = 0$ is supposed to impose free endpoint conditions. That means, if we try to impose an artificial dependence on some boundary data $\tau_a$, the Mach constraint will be able to use the extra gauge freedom to eliminate this dependence. Indeed, the order of the integrations that we will use will highlight this process explicitly. In the second step, we will see how the gauge freedom introduced by the linear momentum constraint $\calf$ is sufficient to eliminate the kernel's dependence on $\tau_q$. This is how the corrected coordinate method is successful at eliminating time from the quantum mechanical theory. Even though this process occurs in exactly the same manner as the elimination of the $\tau_a$ dependence, we will use an alternate order for doing the integrals that will highlight the cancelation of the $\tau_q$ dependence explicitly.

\begin{description}
  \item[Step 1] First impose the Mach constraint by integrating over $dM_\alpha$ for $\alpha = 0, \hdots, N-1$. This will 	 produce the $N$ $\delta$-functions $\delta(\pi^\alpha)$. Integrating over $d\pi^\alpha$ will then impose the Mach constraint
	\equa{\pi^\alpha = 0.}
	Once we have done this we can use the gauge fixing functions $h_\alpha$ to fix a gauge for the $\dota$'s. To do this we will integrate over $dS^\alpha$ for $\alpha = 1, \hdots, N-1$. Note that here $\alpha$ ranges from 1 to $N-1$ only. That is, we will still have an important integration over $dS^0$. Holding off the $dS^0$ integration for now will give the $N-1$ $\delta$-functions $\delta(\dota_\alpha - h_\alpha)$ which imply \equa{\dota_\alpha = h_\alpha} after integrating over $d\dota_\alpha$.

      After setting all the $\pi$'s to zero and the $\dota$'s to $h$'s, we can complete the final integrations over $dS^0$ and $d\dota_0$. The freedom to vary $\dota_0$ arbitrarily is unique to the free endpoint variation. This fact is crucial to the disappearance of the $\tau_a$ dependence of the theory. The remaining terms proportional to $S^0$ are
      \equa{
        i S^0( \tau_a - \Delta\lambda_0 \dota_0 - \sum_{\alpha = 1}^{N-1} \Delta\lambda_\alpha h_\alpha )
      }
      telling us that the $dS^0$ and $d\dota_0$ integrations will impose the condition
      \equa{
         \Delta\lambda_0 \dota_0 = \tau_a - \sum_{\alpha = 1}^{N-1} \Delta\lambda_\alpha h_\alpha.
      }
      which should be solved for $\dota_0$. But, since there are no further $\dota_0$ terms because of the disappearance of the $\pi$'s due to the Mach constraint, this is just an interesting but irrelevant fact that no longer affects any of the remaining terms in the kernel. We can see that the freedom to vary the endpoints of the $a$'s coupled with gauge freedom of the $h$'s gives us a theory that, at the end of the day, doesn't depend on the $\tau_a$'s.

  \item[Step 2] To see how time is eliminated from the quantum theory we could proceed with the same order of integration as before but substitute the Mach constraint with the linear momentum constraint $\calf$ and the $\dotq^0$'s for the $\dota$'s. This would lead to the irrelevant condition
       \equa{
         \Delta\lambda_0 \dotq^0_0 = \tau_q - \sum_{\alpha = 1}^{N-1} \Delta\lambda_\alpha g_\alpha.
      }
      for\footnote{See \apx{A} Step 2 for more details of this calculation.} $\dotq^0_0$. However, it is instructive switch the order of the integrations. In this case, we still impose the linear momentum constraint first by integrating over $dL_\alpha$ followed by $dp^\alpha_0$. This should be done for \emph{all} $\alpha$'s and will force the vanishing of the energy: \equa{p^\alpha_0 = 0.} This will kill the $\dotq^0_\alpha p^\alpha_0$ terms leaving only the $-R^\alpha \dotq^0_\alpha$ terms proportional to $\dotq^0_\alpha$. Thus, the linear momentum constraint has left us with only a single $\dotq^0_\alpha$ term so if we integrate over $\dotq^0$ \emph{first} then $R^\alpha$ we will get \equa{R^\alpha = 0.} For $\alpha = 0$ this kills the $\tau_q$ term and many of the $g_\alpha$ terms. For $\alpha \neq 0$, this kills the rest of the $g_\alpha$ terms. This method leaves no doubt that the extra gauge freedom destroys the time dependence of the kernel. The previous method, however, gives us better physical intuition for exactly how the gauge freedom accomplishes this.
\end{description}

After these integrations, we are left with the kernel
\begin{multline}
    K(Q^i_I, \tau_q,\tau_a) = K_{\jbb}(Q^i_I) = \int \cald q^i_I \cald \Eps \cald p^I_i \cald N \abs{\pb{\ham}{\calg}}
    \times \exp \lf\{ i \int d\lambda \lf[ \sum_I \dotq^i_I p^{I}_i -\rt.\rt. \\
    \lf. \lf. N \lf( V + \sum_I \frac{p_{I}^2}{2m_I} \rt) - \Eps \lf( \sum_I \frac{m_I \dotq^i_I p^I_i}{p_I^2} - f \rt) \rt] \rt\}.
\end{multline}
This is identical to the kernel for JBB theory derived in \cite{sg:emer_time_PI} generalized to many particles.

\subsection{Fixed Endpoint Time-Gauged PPM} \label{sec:fixedpnm}

There are two major differences between the \emph{fixed} endpoint variation and the \emph{free} endpoint variation. The first is that we no longer have the Mach constraint. The mathematical result of this is simply to remove two terms from the Lagrangian: the term proportional to the Mach constraint itself and the term proportion to its gauge fixing constraint. The second major difference is in how the boundary conditions are applied. Because we no longer have gauge fixing functions to fix the values of the auxiliary fields we must specify these the old fashioned way on the boundary. For the moment, this will mean that our kernel will depend on $\tau_a$ and $\tau_q$ though we will see later that, in fact, it will only depend on the combination $\tau_a + \tau_q$ which can be clumped into a single time variable $\tau$. It is instructive to go through the same steps as before to see precisely how the fixed endpoint variation will reintroduce a time dependence into the theory.

\subsubsection{Evaluating the Path Integral}

We can impose the boundary conditions the old fashioned way by removing the integral over $\dota_0$ and insert instead its value
\equa{\label{eq:a0val}
    \Delta\lambda_0 \dota_0 = \tau_a - \sum_{\alpha = 1}^{N-1} \lambda_\alpha \dota_\alpha
}
in terms of $\dota_\alpha$ and $\tau_\alpha$. Inserting this into the action we start with kernel
\begin{multline} \label{eq:fixedkernel}
    K(Q^i_I, \tau_q,\tau_a) = \int_{-\infty}^\infty \frac{d^d p_i^{I,0}}{(2\pi)^d} \frac{dN_0}{2\pi} d \dotq^0_{0} \frac{dR^0}{2\pi} \frac{d p_0^{0}}{2\pi} \frac{dL_0}{2\pi} \frac{d \pi^{0}}{2\pi} \times \exp \lf\{ i \lf( \pi_0\tau_a + R_0\tau_q \rt) \rt\} \\
    \times \exp \lf\{ i \Delta\lambda_0 \lf[ \sum_I \dotq^i_{I,0} p^{I,0}_i + \dotq^0_0 p_0^0  - N_0\lf( p_0^0 + V^0 + \sum_I \frac{p_{I,0}^2}{2m_I} \rt) - L_0(p_0^0 - \pi^0) - R^0 \dotq^0_0 \rt] \rt\} \\ \prod_{\alpha = 1}^{N-1} d^d q^i_{I,\alpha} \frac{d\Eps^\alpha}{2\pi} \frac{d^d p_i^{I,\alpha}}{(2\pi)^d} \frac{dN_\alpha}{2\pi} d \dotq^0_{\alpha} \frac{dR^\alpha}{2\pi} \frac{d p_0^{\alpha}}{2\pi} \frac{dL_\alpha}{2\pi} d \dota_{\alpha} \frac{d \pi^{\alpha}}{2\pi} \\
    \times \exp \lf\{ i \Delta\lambda_\alpha \lf[ \sum_I \dotq^i_{I,\alpha} p^{I,\alpha}_i + \dotq^0_\alpha p_0^\alpha + \dota_\alpha (\pi^\alpha - \pi_0) - N_\alpha\lf( p_0^\alpha + V^\alpha + \sum_I \frac{p_{I,\alpha}^2}{2m_I} \rt) - \rt.\rt. \\
    \lf. \lf. \Eps^\alpha \lf( \sum_I \frac{m_I \dotq^i_{I,\alpha} p^{I,\alpha}_i}{p_{I,\alpha}^2} - f_\alpha \rt) - L_\alpha(p_0^\alpha - \pi^\alpha) - R^\alpha (\dotq^0_\alpha - g_\alpha) -R^0 g_\alpha \rt] \rt\}\times \abs{\pb{\ham}{\calg}}.
\end{multline}
Notable differences between this kernel and the one of \eq{longkernel} include the disappearance of the $M$ and $S$ terms as well as the $d\dota_0$ integration. Instead we have a new $\dota_\alpha \pi_0$ term and $S_0$ has been replaced by $\pi_0$ in $e^{iS_0\tau_a}\ra e^{i\pi_0 \tau_a}$. The next step is to perform the integrations in the same way we did for the free endpoint variation.

\begin{description}
    \item[Step 1] We can no longer impose the Mach constraint, of course, as it no longer exists in the problem nor can we fix a gauge for the $a$'s. We can, however, integrate over the remaining variables needed to bring us to the beginning of Step 2. These are just the $\dota_\alpha$'s and the $\pi^\alpha$'s. Because we have no $\dota_0$ integration, we will just keep the $\pi_0$ integration for now as it will ultimately correspond to an integration over all energies later. Integrating over the $\dota_\alpha$'s first will give
        \equa{
            \pi^\alpha = \pi^0
        }
        which leaves us with a kernel of the form
        \begin{equation}
            K(Q^i_I, \tau_q,\tau_a)_{\text{Step 1, fixed}} = \int \frac{d\pi_0}{2\pi} e^{i\pi_0\tau_a} \int \cald\mu\, I(Q^i_I, \tau_q)_{\text{Step 1, free}}\exp \lf\{ i \sum_{\alpha = 0}^{N-1} \Delta\lambda_\alpha L_\alpha\pi^0\rt\}
        \end{equation}
        where $\cald\mu$ is short for the integration measure of \eq{kernelmaster} and $I(Q^i_I, \tau_q)_{\text{Step 1, free}}$ is its integrand after Step 1. That is, we can write the result for the kernel after Step 1 of the \emph{fixed} endpoint variation, $K(Q^i_I, \tau_q,\tau_a)_{\text{Step 1, fixed}}$, explicitly as an integral over the integrand after Step 1 of the \emph{free} endpoint variation, $I(Q^i_I, \tau_q)_{\text{Step 1, free}}$ times some simple exponentials. Note that the $\tau_a$ dependence has been restored.

    \item[Step 2] In Step 2, we can proceed exactly as before with our results being only slightly different. Imposing the linear momentum constraint by integrating first over $dL_\alpha$ then $\pi_\alpha$ will give
        \equa{
            p^\alpha_0 = \pi^0.
        }
        This can be done for all values of $\alpha$ including $0$. It is why the integration over $\pi^0$ will effectively give an integration over an energy instead of the previous case where $p^\alpha$ was just 0.

        By integrating over the remaining variables we will see what happens to the $\tau_q$ dependence. First, integrate over $dR^\alpha$ for $\alpha \neq 0$ then integrate over $d\dotq^0_\alpha$. This will set
        \equa{
            \dotq^0_\alpha = g_\alpha.
        }
        Finally, if we integrate over $dR_0$ then $d\dotq^0_0$ we get
        \equa{
            \Delta\lambda_0 \dotq^0_0 = \tau_q - \sum_{\alpha = 1}^{N-1} \Delta\lambda_\alpha g_\alpha
        }
        as an expression for $\dotq^0_0$. This, however, is no longer an irrelevant fact as we still have a remaining $\dotq^0$ term proportional to $\pi^0$. Inserting this into our remaining integral and noting that the $g_\alpha$ terms cancel, as they should, in the final kernel, we obtain
        \begin{multline}
              K(Q^i_I, \tau_q,\tau_a) = \int \frac{d\pi^0}{2\pi} e^{\pi^0(\tau_a + \tau_q)} \int \cald q^i_I \cald \Eps \cald p^I_i \cald N \abs{\pb{\ham}{\calg}}\\
               \times \exp \lf\{ i \int d\lambda \lf[ \sum_I \dotq^i_I p^{I}_i - N \lf( \pi^0 + V + \sum_I \frac{p_{I}^2}{2m_I} \rt) - \Eps \lf( \sum_I \frac{m_I \dotq^i_I p^I_i}{p_I^2} - f \rt) \rt] \rt\}.
        \end{multline}
        Calling $\pi^0 = -E$ and $\tau = \tau_a + \tau_q$ we get
        \equa{
            K(Q^i_I, \tau)_{\text{fixed}} = \int \frac{dE}{2\pi} e^{iE\tau} K_{\jbb}(Q^i_I, E)
        }
        which is exactly the expression for the kernel of PPM derived in \cite{sg:emer_time_PI}. It is effectively the Fourier transform of the JBB kernel for energy eigenstates evaluated at a particular time $\tau$. Note that $\tau$ is shifted from what we would expect by trying to apply the boundary condition $\tau_q$. One can simply understand this as the result of having an absolute time defined by the absolute orientation of the $a$'s. This kernel is a solution to the time \emph{dependant} Schr\"{o}dinger equation. Hence, we have demonstrated that fixing the endpoints of the variation of the auxiliary fields reinserts a background structure into the theory from which a time dependence emerges.
\end{description}

\section{Results / Conclusions}

\subsection{Summary of Results}

Our first main result is that time-gauged PPM is equivalent classically and quantum mechanically to JBB theory. More specifically, we showed that the background absolute time structure present in PPM can be removed by applying the corrected coordinate method of Barbour to its time translational invariance. Removing this background structure completely removed the time dependence of the quantum theory. In the canonical approach, the disappearance of time was a result of the Mach constraint killing the energy operator and demoting the time dependent \schro equation to the time independent \schro equation. In the path integral approach, time disappeared because the extra gauge freedom provided by the Mach constraint allowed us to absorb the kernel's dependence on any boundary conditions associated with time into non-physical degrees of freedom. In both cases, we found that introducing the auxiliary fields $a$ and allowing the endpoints to vary freely was sufficient to eliminate the background absolute structure and render the theory temporally relational.

Our second key result concerned returning the absolute structure to time-gauged PPM by altering the corrected coordinate method to fix the endpoints of the auxiliary fields. More precisely, we found that fixing the endpoints of the $a$'s produced a theory that is classically and quantum mechanically equivalent to PPM. In the canonical framework, we can see the background structure emerge as a result of the disappearance of the Mach constraint leaving us with a time dependent \schro equation. In the path integral framework, fixing the endpoints means we don't have enough gauge freedom to absorb the time dependence into non-physical degrees of freedom leaving a true physical dependence on the boundary data.

Collecting these results, we obtain a specific criterion for distinguishing a general reparameterization invariant theory from a a truly relational theory: if there is a symmetry in the action, a \emph{free} endpoint variation of the auxiliary fields associated to that symmetry will generate a relational theory while a \emph{fixed} endpoint variation will restore the background structure.

\subsection{Conclusions / Open Questions}

What conclusions can we draw from these results? If we take best-matching or the corrected-coordinate method (which has been shown to be intimately connected to the gauge principle in \cite{gryb:ym_bm}) as a fundamental principle of Nature then we are led immediately to two choices: 1) we keep the endpoints of the auxiliary fields fixed as if they were standard physical variables, or 2) we allow the endpoints to vary freely admitting no knowledge of their absolute orientation. Note that only the second option properly implements Machian ideas while the first option assumes the existence of absolute structures.

According to the analysis in this paper, choosing option 2) would lead to a proper relational theory \emph{even} if the variable we are making relational is time. This fact provides an alternative to using Jacobi's principle for relational time proposed by Barbour. Admittedly, this approach is not as nice conceptually since there is no obvious way to interpret the action principle as a geodesic principle on configuration space. However, we have shown that it is mathematically equivalent to Jacobi's principle not only classically but also quantum mechanically. In \apx{spatial_sym}, we show that these results are not modified by best-matching the spatial symmetries. Thus, as an alternative to Jacobi's principle, it may not be as conceptually pleasing but, owing to the lack of the square root, it is mathematically cleaner. Furthermore, if our results regarding the free endpoint variation extend to geometrodynamics, as we might expect based on the link between GR and our toy model discussed in the introduction, then the ADM action may have less in common with the action of PPM, as has been often suggested \cite{adm:adm_review,kuchar:prob_of_time}, and more in common with the action of time-gauged PPM or, alternatively, the JBB action. If the variation carried out by ADM is equivalent to a free endpoint variation of the auxiliary fields of time gauged geometrodynamics then one should expect this theory to be temporally relational and suffer from a Problem of Time. The fact that all efforts to study the ADM action have led to a Problem of Time would suggest that this is indeed the case.

Alternatively, we can choose option 1). Our analysis in this paper would suggest that such a choice should lead to a theory with a definite background structure. This raises two interesting questions:

First, is there a way to introduce a background structure into GR? This is not a new question and has been raised most notably by Kucha$\check{\text{r}}$ after noting the similarities between the ADM action and the action of PPM \cite{kuchar:lect_notes, kuchar:prob_of_time}. He conjectured that GR might have a hidden background structure and that solving the Problem of Time would involve finding a way to write the theory in terms of this background structure and quantize it. In light of our results, we can understand why this has yet to be achieved\footnote{For alternative explanations of why these attempts may have failed, see \cite{torre:gr_param}.}. If, as in option 2), the variation of the ADM is equivalent to a free endpoint variation then the theory will be relational. On the other hand, if we choose option 1) we might be able to introduce a background structure in GR in the same way we have done for time gauged PPM. Implementing this choice would involve removing the Mach constraint from time gauged geometrodynamics. We are currently exploring this option and its connection to the ADM action. Indeed, it may be that implementing option 1) and option 2) would lead respectively to the time independent and time dependent Wheeler-DeWitt equations considered by York and Brown \cite{brown:gr_time}. It appears that the main difficulty in deriving the theories of York and Brown using the corrected coordinate method on the dynamic 3-geometry analogue of PPM is in motivating the use of a time parameter that is local in space\footnote{The presence of a \emph{local} time variable is the analogue of the \emph{local} square root in the BSW formulation.}. This is the only significant difference, other than the choice of configuration space, between these toy models and GR.

We are now brought to our second question: what would such a background structure mean in GR? Indeed, what does this background structure even mean in the context of time-gauged PPM? At this point, one could only speculate that this structure could be equivalent to the emergent absolute structures that seem to emerge from relational theories on isolated subsystems of the universe. The absolute frame of rest provided by the fixed stars is a good example of how absolute structures can emerge classically from relational theories. In this case, the local absolute theory is very well approximated by the \emph{fixed} endpoint variation of time-gauged PPM (on and off shell) while the global relational theory is given by the \emph{free} endpoint variation\footnote{One could be more rigorous and consider the global theory to be GR of which time gauged PPM would provide the weak gravity, non-relativistic limit.}. For non-relativistic particle mechanics, the emergence of absolute structures on isolated subsystems can be understood classically \cite{barbour:mach_principle} but quantum mechanically we are still lacking a convincing demonstration of this. The path integral techniques developed in this paper and in \cite{sg:emer_time_PI} are ideally suited to studying this problem. In principle, it should be possible to extend these techniques to geometrodynamics in order to study the Problem of Time in quantum GR.

Can we introduce an absolute structure into geometrodynamics, in accordance with \cite{brown:gr_time}, by varying the auxiliary fields of time gauged geometrodynamics while fixing their endpoints? Would the theory we obtain in this way be emergent from GR, a theory which is fully relational, on isolated subsystems of the universe? These are all interesting open questions that can be answered if the results of this work do indeed extend to geometrodynamics.

\section{Acknowledgements}

I would like to warmly thank Julian Barbour for inviting me to the beautiful College Farm, where I learned much about Mach's principle, and for contributing very helpful comments. I would also like to thank Ed Anderson and Brenden Foster for useful comments on the draft as well as Lee Smolin for advice and guidance. Research at the Perimeter Institute is supported in part by the Government of Canada through NSERC and by the Province of Ontario through MEDT. I also acknowledge support from an NSERC Postgraduate Scholarship, Mini-Grant MGA-08-008 from the Foundational Questions Institute (fqxi.org), and from the University of Waterloo.

\appendix

\section{Spatial Symmetries}\label{apx:spatial_sym}

In the full BB theory, the spatial symmetries (ie, translations, rotations, and scale invariance) of classical mechanics are best-matched. This introduces linear constraints on the generalized momenta which are analogous to the diffeomorphism constraints of GR \cite{barbourbertotti:mach,barbour:scale_inv_particles,barbour_el_al:scale_inv_gravity}. Thus, to be truly analogous to GR we should consider the spatial symmetries as well. For example, one might worry that the non-commutativity of the rotation group might ruin the algebraic structure that allows for the equivalence between the JBB theory and time-gauged PPM. In this section we will add the spatial symmetries to time-gauged PPM and compare this to the full BB theory. To prove the equivalence, it will be sufficient to study the Hamiltonian formulation. We will use the modified CCM introduced in \cite{gryb:ym_bm}.

\subsection{Momenta and Constraints}

Following the procedure for the modified CCM outlined in \cite{gryb:ym_bm}, we start with the actions (\ref{eq:jbb_action}) of JBB theory and the action (\ref{eq:action}) for time-gauged PPM then substitute derivatives with respect to $\lambda$ for covariant derivatives with a flat connection. That is, we make the substitutions
\equa{
    \dot{q}^i_I \ra \dot{q}^i_I + \dot{\omega}^\alpha \lf. t_\alpha\rt.^i_j q^j_I,
}
where the $\omega^\alpha$'s are the auxiliary fields which are group parameters associated with a particular spatial symmetry group generated by $t_\alpha$. They are varied using a \emph{free} endpoint variation using a Mach constraint. In what follows, we will suppress spatial indices and use a matrix notion. Row vectors will be denoted with a transpose $^T$ and column vectors will appear with no indices. $t_\alpha$ is the only matrix we will need.

With this notation, the new actions $S_\text{BB}$ for BB theory and $S_\text{BMPPM}$ for Best-Matched PPM become
\begin{align}
    S_\text{BB} &= \int d\lambda\, 2\, \sqrt{T(q_I + \dot{\omega}^\alpha t_\alpha q_I)} \sqrt{V(q_I)}, \quad \text{and} \\
    S_\text{BMPPM} &= \int d\lambda\, \lf[ \frac{T(q_I + \dot{\omega}^\alpha t_\alpha q_I)}{\dot{q}^0 + \dot{a}} - (\dot{q}^0 + \dot{a})V(q_I) \rt].
\end{align}

We can now compute the definitions of the momenta, $\pIt$, conjugate to $\qI$ and the momenta, $\pia$, conjugate to $\omegaa$. For BB theory these are:
\begin{align}
    \pIt &= \diby{L_\text{BB}}{\dotq_I} = \sqrt{\frac{E-V}{T}} m_I (\dotq_I + \dot{\omega}^\alpha \ta \qI)^T, \quad \text{and} \\
    \pia &= \diby{L_\text{BB}}{\dotomega^\alpha} = \sqrt{\frac{E-V}{T}} \sum_I m_I(\dotq_I + \dot{\omega}^\alpha \ta \qI)^T \ta \qI.
\end{align}
For BMPPM, these are:
\begin{align}
    \pIt &= \diby{L_{\text{BMPPM}}}{\dotq_I} = \frac{1}{\dotq^0 + \dota} m_I(\dotq_I + \dot{\omega}^\alpha \ta \qI)^T, \quad \text{and} \\
    \pia &= \diby{L_{\text{BMPPM}}}{\dotomega^\alpha} = \frac{1}{\dotq^0 + \dota} \sum_I m_I(\dotq_I + \dot{\omega}^\alpha \ta \qI)^T \ta \qI.
\end{align}
Note that the difference between the definitions of the momenta in BB theory and in BMPPM is the replacement
\equa{
    \sqrt{\frac{E-V}{T}} \ra \frac{1}{\dotq^0 + \dota}.\label{eq:subs}
}
For BMPPM, we have the additional momenta $p_0$ and $\pi_0$ conjugate to $q^0$ and $a$ whose expressions are unchanged from \eq{ppm_p0} and \eq{ppm_pi} other than the fact that $T$ is evaluated at $(\dotq_I + \dot{\omega}^\alpha \ta \qI)$ rather than $\dotq_I$.

In both BB theory and BMPPM, we cannot invert the definitions of the momenta uniquely for the $\dotq$'s and $\dot{\omega}$'s. This indicates the presence of new constraints. Interestingly, because of the simple substitution (\ref{eq:subs}), these constraints take the same form for both theories. We then define the new constraints $\lin_\alpha$
\equa{
    \lin_\alpha = \pia - \sum_I \pIt \ta \qI = 0,
}
which are linear in the momenta and should be satisfied by both theories. To impose the free endpoint condition, we will also need the Mach constraints
\equa{
    \pia = 0.
}
Finally, we have the quadratic scalar constraints
\begin{align}
    \ham_{\text{BB}} &= V + \sum_I \frac{(p^I_i)^2}{2m_I} - E = 0 \quad \text{and} \\
    \ham_{\text{BMPPM}} &= V + \sum_I \frac{(p^I_i)^2}{2m_I} + p_0 = 0,
\end{align}
which are unchanged by the presence of the spatial symmetries. In BMPPM, we also have the additional linear constraint
\equa{
    \lin_0 = p_0 - \pi = 0
}
and the Mach constraint
\equa{
    \pi_0 = 0.
}

It is easy to show that these two systems of constraints are equivalent. If we substitute $\pi_0 = 0$ into $\lin_0 = 0$ we find that $p_0 = 0$. Inserting this into $\ham_{\text{BMPPM}}$ gives $\ham_{\text{BB}}$ provided we absorb the total energy $E$ back into the potential $V$. The remaining constraint are trivially identical.

\subsection{Constraint Algebra and Equations of Motion}

The presence of the new phase space variables $\omegaa$ and $\pia$ imply the new fundamental Poisson Brackets (PB)
\equa{
    \pb{\omegaa}{\pi_\beta} = \delta^\alpha_\beta.
}
A moment's inspection will reveal that the only non-trivial PBs introduced by the new constraints are
\equa{
    \pb{\lin_\alpha}{\ham_\text{BB,BMPPM}}.
}
For both theories, this will be zero provided
\equa{
    \sum_I\lf[ \frac{1}{m_I} \pIt \ta p_I - \diby{V}{\qI}\ta \qI \rt] = 0. \label{eq:ccs}
}
This is a requirement of the gauge invariance of the original action. From the perspective of BB theory, it is a consistency requirement on the form of the potential \cite{barbour:scale_inv_particles}. Thus, for potentials satisfying (\ref{eq:ccs}), the constraint algebra is first class and we have no secondary constraints.

It is easy to show that the canonical Hamiltonian for both systems is identically equal to zero as it should be in any reparameterization invariant theory. The total Hamiltonians for both theories are given by:
\begin{align}
    H_\text{T,BB} &= L^\alpha\lin_\alpha + M^\alpha\pia + N\,\ham_\text{BB}, \quad \text{and} \\
    H_\text{T,BMPPM} &= L^\alpha\lin_\alpha + M^\alpha\pia + N\,\ham_\text{BMPPM} + M^0\pi_0 + L^0\lin_0.
\end{align}
Just as in the case with no spatial symmetries, the $\pi_0$ and $\lin_0$ terms are used to make the $\ham_\text{BMPPM}$ term equivalent to the $\ham_\text{BB}$ term. It is clear that the effect of the spatial symmetries is to impose the additional constraints $\lin_\alpha$ and $\pia$ which does not affect any of the discussions regarding time.

As a final check, we can compute the classical equations of motion. For both theories we have,
\begin{align}
    \dotq_I &= \pb{\qI}{H_\text{T}} = N \frac{p_I}{m_I} - L^\alpha \ta \qI,  \\
    \dotp_I^T &= \pb{\pIt}{H_\text{T}} = -N \diby{V}{\qI} + L^\alpha \pIt \ta, \\
    \dot{\omega}^\alpha &= \pb{\omegaa}{H_\text{T}} = L^\alpha + M^\alpha, \quad \text{and} \\
    \dot{\pi}_\alpha &= \pb{\pia}{H_\text{T}} = 0.
\end{align}
These are just the Hamiltonian form of the equations of motion of standard BB theory \cite{barbour:scale_inv_particles}. The final equations of motion of PPM are
\begin{align}
    \dot{q}^0 &= \pb{q^0}{H_\text{T}} = N + L^0,\\
    \dotp_0 &= \pb{p_0}{H_\text{T}} = 0,\\
    \dota &= \pb{a}{H_\text{T}} = M^0 - L^0, \quad \text{and} \\
    \dot{\pi} &= \pb{\pi}{H_\text{T}} = 0.
\end{align}
They are completely unchanged by the presence of the spatial symmetries. Thus, their affect will be no different from that discussed in \scn{eom}. We can then be sure that adding spatial symmetries won't affect the discussions regarding time that have been considered in this paper even if the symmetry groups in question are non-commutative.

\section{Free Endpoint Path Integral With No Mach constraint} \label{apx:A}

In this appendix we will recalculate the kernel evaluated in \scn{freepnm} but using a technique that does not involve the use of a Mach constraint. In the canonical formulation of PPM, the Mach constraint is responsible for guaranteeing that the variation of the auxiliary field is free at the endpoints. In the path integral, it is possible to achieve the same thing without using a Mach constraint. The free endpoint variation can be implemented by using the fixed endpoint kernel of \eq{fixedkernel} and integrating over all values of $\dota_0$ rather then inserting the specific value given by (\ref{eq:a0val}). This gives an additional integration over $d\dota_0$ and a term $\Delta\lambda_0\pi^0\dota_0$ rather then $ \pi^0( \tau_a - \sum_{\alpha = 1}^{N-1} \Delta\lambda_\alpha\dota_\alpha)$. Our starting point is, then, the kernel
\begin{multline}
    K(Q^i_I, \tau_q) = \int_{-\infty}^\infty \frac{d^d p_i^{I,0}}{(2\pi)^d} \frac{dN_0}{2\pi} d \dotq^0_{0} \frac{dR^0}{2\pi} \frac{d p_0^{0}}{2\pi} \frac{dL_0}{2\pi} d\dota_0 \frac{d \pi^{0}}{2\pi} \times \exp \lf\{ i R_0\tau_q \rt\} \\
    \times \exp \lf\{ i \Delta\lambda_0 \lf[ \sum_I \dotq^i_{I,0} p^{I,0}_i + \dotq^0_0 p_0^0 + \dota_0 \pi^0 - N_0\lf( p_0^0 + V^0 + \sum_I \frac{p_{I,0}^2}{2m_I} \rt) - L_0(p_0^0 - \pi^0) - R^0 \dotq^0_0 \rt] \rt\} \\ \prod_{\alpha = 1}^{N-1} d^d q^i_{I,\alpha} \frac{d\Eps^\alpha}{2\pi} \frac{d^d p_i^{I,\alpha}}{(2\pi)^d} \frac{dN_\alpha}{2\pi} d \dotq^0_{\alpha} \frac{dR^\alpha}{2\pi} \frac{d p_0^{\alpha}}{2\pi} \frac{dL_\alpha}{2\pi} d \dota_{\alpha} \frac{d \pi^{\alpha}}{2\pi} \\
    \times \exp \lf\{ i \Delta\lambda_\alpha \lf[ \sum_I \dotq^i_{I,\alpha} p^{I,\alpha}_i + \dotq^0_\alpha p_0^\alpha + \dota_\alpha \pi^\alpha - N_\alpha\lf( p_0^\alpha + V^\alpha + \sum_I \frac{p_{I,\alpha}^2}{2m_I} \rt) - \rt.\rt. \\
    \lf. \lf. \Eps^\alpha \lf( \sum_I \frac{m_I \dotq^i_{I,\alpha} p^{I,\alpha}_i}{p_{I,\alpha}^2} - f_\alpha \rt) - L_\alpha(p_0^\alpha - \pi^\alpha) - R^\alpha (\dotq^0_\alpha - g_\alpha) -R^0 g_\alpha \rt] \rt\}\times \abs{\pb{\ham}{\calg}}.
\end{multline}

\begin{description}
    \item[Step 1] We need to integrate over the $\dota_\alpha$'s then the $\pi^\alpha$'s . It is now trivial to do this for $\alpha = 0\hdots (N-1)$. The result is simply to force the vanishing of the $\pi_\alpha$'s
        \equa{\label{eq:vanshingp}
            \pi^\alpha = 0.
        }
        In the method described in \scn{freepnm}, the vanishing of these momenta is the result of imposing the Mach constraint. Here, this is seen as arising from the fact that the Hamiltonian does not contain any $\dota$ terms meaning that the Lagrangian only has one proportional to $\pi$. Applying the boundary conditions, as in \scn{fixedpnm}, introduces additional $\dota$ terms to the Hamiltonian which, in turn, contribute additional terms to the RHS of \eq{vanshingp}. This is the main difference between the two types of variation.
    \item[Step 2] This step is identical to Step 2 of \scn{evalfree}. Here, we will use the alternative method discussed in that section to obtain the final result. First, we impose the linear momentum constraint by integrating over $dL_\alpha$ then $d\pi^\alpha$ for all $\alpha$'s. This will imply the vanishing of the energy,
        \equa{
            p_0^\alpha = 0,
        }
        as expected for a Machian theory. Integrating over $dR^\alpha$ then $d\dotq^0_\alpha$ for $\alpha = 1\hdots (N-1)$ sets $\dotq^0_\alpha = g_\alpha$ setting a time gauge. The last two integrations over $dR^0$ then $d\dotq^0$ will give the requirement
        \equa{
             \Delta\lambda_0 \dotq^0_0 = \tau_q - \sum_{\alpha = 1}^{N-1} \Delta\lambda_\alpha g_\alpha
        }
        which does not affect the remainder of the kernel.
\end{description}

After these integrations, we are left once again with the kernel
\equa{
        K(Q^i_I, \tau_q,\tau_a) = K_{\jbb}(Q^i_I)
}
agreeing with the results from \scn{freepnm}.

This alternative method differs from the previous one in that it is simpler to implement mathematically and is closer to the method used in the fixed endpoint variation. The difference is that the vanishing of the energy is accomplished by integrating over all the $\dota$'s and having a Hamiltonian independent of the $\dota$'s. This is not as enlightening as the Mach constraint method where the gauge freedom of the Mach constraint is used to eliminate the dependence of the theory on the boundary data. In addition, the Mach constraint method is closer to and inspired by the canonical quantization of the theory. However, we note in passing that, from the path integral perspective, the Mach constraint or, more specifically, the associated Machian invariance, though still an invariance of the theory, brings with it extra baggage that is not necessary for actually calculating the kernel. In a sense, the free endpoint variation is a stronger criterion for relationalism than the Mach constraint.

% ---------------------------------------- Bibliography ------------------------------------------------

\bibliographystyle{unsrt}
\bibliography{mach}

\end{document}